\documentclass[conference]{IEEEtran}
\ifCLASSINFOpdf
  \usepackage[pdftex]{graphicx}
\else
\fi
\usepackage[cmex10]{amsmath}
\usepackage{lettrine}
\usepackage[linesnumbered,ruled,vlined]{algorithm2e}
\usepackage{amssymb}
\usepackage{bm}
\usepackage{xcolor}
\usepackage[nameinlink]{cleveref}
\crefname{figure}{Fig.}{Figs.}
\begin{document}
%
\title{UAV-Assisted Enhanced Coverage and Capacity in Dynamic MU-mMIMO IoT Systems: A Deep Reinforcement Learning Approach\vspace{-12mm}}
\author{\IEEEauthorblockN{MohammadMahdi Ghadaksaz, Mobeen Mahmood, Tho Le-Ngoc}
\IEEEauthorblockA{Department of Electrical and Computer Engineering
McGill University, Montreal, QC, Canada\\
Email: mohammad.ghadaksaz@mail.mcgill.ca, mobeen.mahmood@mail.mcgill.ca, tho.le-ngoc@mcgill.ca }}
\maketitle

\begin{abstract}
This study focuses on a multi-user massive multiple-input multiple-output (MU-mMIMO) system by incorporating an unmanned aerial vehicle (UAV) as a decode-and-forward (DF) relay between the base station (BS) and multiple Internet-of-Things (IoT) devices. Our primary objective is to maximize the overall achievable rate (AR) by introducing a novel framework that integrates joint hybrid beamforming (HBF) and UAV localization in dynamic MU-mMIMO IoT systems. Particularly, HBF stages for BS and UAV are designed by leveraging slow time-varying angular information, whereas a deep reinforcement learning (RL) algorithm, namely deep deterministic policy gradient (DDPG) with continuous action space, is developed to train the UAV for its deployment. By using a customized reward function, the RL agent learns an optimal UAV deployment policy capable of adapting to both static and dynamic environments. The illustrative results show that the proposed DDPG-based UAV deployment (DDPG-UD) can achieve approximately 99.5\% of the sum-rate capacity achieved by particle swarm optimization (PSO)-based UAV deployment (PSO-UD), while requiring a significantly reduced runtime at approximately 68.50\% of that needed by PSO-UD, offering an efficient solution in dynamic MU-mMIMO environments.
\end{abstract}


%
\IEEEpeerreviewmaketitle
\vspace{-2ex}

\section{Introduction}
\vspace{-1ex}
\lettrine{I} {n} the next-generation of wireless communications, the expectation of connectivity anywhere and anytime poses a formidable challenge, particularly in dynamic Internet-of-Things (IoT) environments where the users are continuously on the move, changing their locations frequently. These IoT environments, spanning from ever-changing urban areas to critical emergencies, necessitate a network infrastructure that is both adaptable and robust. In this case, several networking strategies have been explored, such as direct transmission and relay-based configurations. However, direct transmission over large distances can be impractical and result in excessive power consumption. It also frequently fails to fulfill the dynamic demands and changing conditions of these areas when employing the traditional static structure of base station (BS). Under these conditions, employing mobile relay nodes emerges as a more energy-efficient solution \cite{ref:1}.  \par
Recent developments in unmanned aerial vehicles (UAVs), commonly referred to as drones, have positioned them as an essential element of the future wireless communications networks. When used as relays, UAVs offer several advantages over traditional static relay systems. Specifically, mobile, on-demand relay systems are highly suitable for unforeseen or transient events, like emergencies or network offloading tasks, due to their ability to be quickly and economically deployed \cite{ref:2}. The mobility of UAVs enables them to operate at relatively high elevations, set up a line-of-sight connection with users on the ground, and prevent signal interference caused by obstacles, which makes UAVs practically appealing for dynamic communications systems \cite{ref:3}. Despite the significant propagation challenges faced by millimeter-wave (mmWave) signals, including free-space path loss, atmospheric and molecular absorption, and attenuation from rain, their substantial bandwidth presents a promising solution for meeting the high-throughput and low-latency requirements of diverse UAV application scenarios \cite{ref:4}. To address these challenges, massive multiple-input multiple-output (mMIMO) technology is employed, utilizing large antenna arrays to generate robust beam signals and thereby extending the transmission range. Compared to fully-digital beamforming (FDBF), the hybrid beamforming (HBF) architecture, which consists of a radio frequency (RF) stage and a baseband (BB) stage, can minimize power consumption by reducing the number of energy-consuming RF chains while achieving a performance close to FDBF \cite{ref:5}-\cite{ref:7}. \par 
The deployment of the UAV plays an important role in enhancing the performance of UAV-assisted wireless communications systems. Thus, recent research has shown an increased interest in optimizing UAV locations, with particular emphasis on HBF solutions to
maximize the achievable rate (AR) or minimize transmit power \cite{ref:8}-\cite{ref:12}. In particular, \cite{ref:8} investigates the joint optimization of UAV deployment while considering HBF at BS and UAV for maximum AR. An amply-and-forward UAV relay with analog beamforming architecture is considered in \cite{ref:9} to maximize the capacity in a dual-hop mMIMO IoT system. Similarly, \cite{ref:14} considers the optimization problem for UAV location, user clustering, and HBF design to maximize AR under a minimum rate constraint for each user. The authors in \cite{ref:15} study the joint optimization of UAVs flying altitude, position, transmit power, antenna beamwidth, and users’ allocated bandwidth. Most of the existing research works (e.g., \cite{ref:8}-\cite{ref:15}) tackle the issue of UAV deployment in a static environment where users are situated at fixed locations. However, these works tend to overlook the dynamic nature of real-world environments, where ground IoT users/devices exhibit mobility, leading to rapidly changing conditions. \par
As a cornerstone of artificial intelligence (AI), reinforcement learning (RL) has been extensively researched in wireless communications and UAV applications \cite{ref:16}-\cite{ref:18}. RL is a decision-making approach that emphasizes learning through interaction with an environment. Inspired by behavioral psychology, RL is akin to the learning process in humans and animals, where actions are taken based on past experiences and their outcomes. Within the spectrum of RL methods, the deep deterministic policy gradient (DDPG) has emerged as a notable technique \cite{ref:19}. DDPG is particularly adept at handling continuous action spaces, which are common in real-world scenarios. This capability makes DDPG highly relevant for complex, dynamic environments where actions need to be precise and varied, as is often the case with UAV operations. Different RL-based solutions have been studied for UAV positioning (e.g., \cite{ref:20}-\cite{ref:22}). However, the design of HBF jointly with UAV deployment using RL-based solutions is an unaddressed research problem, presenting a significant
opportunity to advance the field of UAV-assisted mMIMO IoT
communications networks in dynamic environments. \par
To address this issue, we propose a joint HBF and UAV deployment framework using the DDPG-based algorithmic solution to maximize AR in dynamic MU-mMIMO IoT systems. In particular, the RF beamforming stages for BS and UAV are designed based on the slow time-varying angle-of-departure (AoD)/angle-of-arrival (AoA) information, and BB stages are formulated using the reduced-dimensional effective channel matrices. Then, a novel DDPG-based algorithmic solution is proposed for UAV deployment with a primary objective to not only maximize the overall AR in MU-mMIMO IoT systems but also to significantly reduce computational complexity, particularly the runtime, compared to nature-inspired (NI) optimization methods. It is worthwhile to mention that the proposed DDPG algorithm exhibits a notable advantage in dynamic scenarios. The knowledge gained from training on the initial user's position is effectively transferred to subsequent positions, resulting in reduced learning time—akin to transfer learning. The illustrative results depict the efficacy of the proposed DDPG-based deployment scheme by reducing the runtime up to 31.5\% as compared to NI-based solutions. \par 
The rest of this paper is organized in the following manner. Section II defines the system and channel model for UAV-relay MU-mMIMO systems. In Section III, we introduce the joint HBF and DDPG-based UAV deployment framework. Section IV presents the illustrative results. Finally, Section V concludes the paper.
\vspace{-1em}
\section{System \& Channel Model} 
\vspace{-1ex}
\subsection{System Model}
\indent The current research explores a complex situation in which various users are linked to a gateway via both wired and wireless connections. This configuration is located in a remote region, which is hard to reach directly by BS because of several hindrances like buildings, mountains, etc. Afterward, a UAV is employed as a dual-hop decode-and-forward (DF) relay to connect with the users as illustrated in Fig. 1. Let ($x_b, y_b, z_b$), ($x_u, y_u, z_u$), and ($x_k, y_k, z_k$) denote the location of the BS, UAV relay, and $k^{th}$ IoT user, respectively. We establish the 3D distances for a UAV-assisted mmWave MU-mMIMO IoT system as follows:
\[\tau_1 = \sqrt{(x_u-x_b)^2+(y_u-y_b)^2+(z_u-z_b)^2}\]
\begin{equation} \label{eq: 1}
\tau_{2, k} = \sqrt{(x_u-x_k)^2+(y_u-y_k)^2+(z_u-z_k)^2}
\end{equation}
\[\tau_k = \sqrt{(x_b-x_k)^2+(y_b-y_k)^2+(z_b-z_k)^2}\]
where $\tau_1$, $\tau_{2, k}$, and $\tau_k$ are the 3D distance between UAV \& BS, between the UAV and $k^{th}$ IoT 
user, and between BS and $k^{th}$ IoT user, respectively. \par
\begin{figure}[!t]
    \centering
    \includegraphics[width=\linewidth]{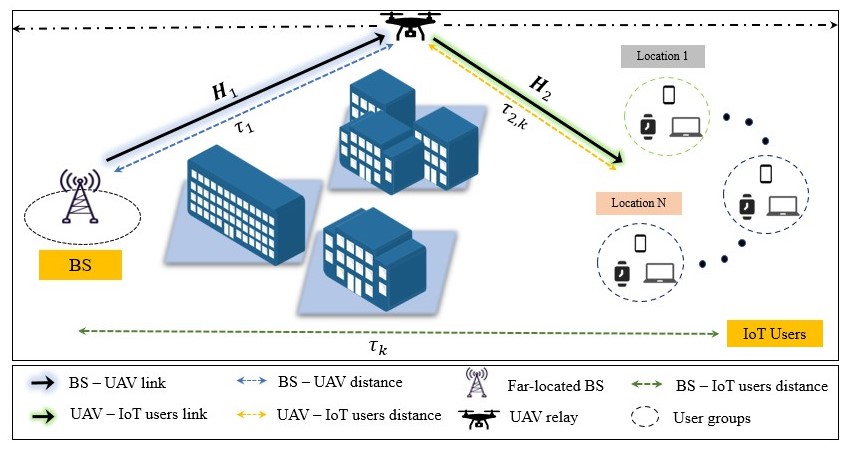}
    \caption{UAV-assisted mmWave MU-mMIMO dynamic environment}
    \label{fig:System Model}
    \vspace{-2ex}
\end{figure}
In this system model, we consider BS equipped with $N_T$ antennas, UAV relay with $N_r$ antennas for receiving and $N_t$ antennas for sending information to $K$ single-antenna IoT users scattered in $G$ groups, where $g^{th}$ group has $K_g$ IoT users such that $K = \sum_{g=1}^G K_g$. Both BS and UAV relay utilize HBF architecture. In this setup, the BS includes an RF beamforming stage $\bold{F}_b\in \mathbb{C}^{N_T\times N_{RF_b}}$ and BB stage $\bold{B}_b\in \mathbb{C}^{N_{RF_b} \times K}$, where $N_{RF_b}$ is the number of RF chains such that $N_s \leq N_{RF_b} \leq N_T$ to guarantee multi-stream transmission. Considering half-duplex (HD) DF relaying, BS sends $K$ data streams $\bold{d} = [d_1, d_2, \cdots, d_K]^T$ through channel $\bold{H}_1 \in \mathbb{C}^{N_r \times N_T}$ in the first time slot. Using $N_r$ antennas, UAV receives signals with RF stage $\bold{F}_{u,r} \in \mathbb{C}^{N_{RF_u} \times N_r}$ and BB stage $\bold{B}_{u,r} \in \mathbb{C}^{K \times N_{RF_u}}$. We assume UAV relay transmits the data in the second time slot using RF beamformer $\bold{F}_{u,t} = [\bold{f}_{u,t,1}\cdots, \bold{f}_{u,t,N_{RF_u}}] \in \mathbb{C}^{N_t \times N_{RF_u}}$ and BB stage $\bold{B}_{u,t} = [\bold{b}_{u,t,1},\cdots,\bold{b}_{u,t,K}] \in \mathbb{C}^{N_{RF_u} \times K}$ via channel $\bold{H}_2 \in \mathbb{C}^{K \times N_t}$. The HBF design significantly cuts down the number of RF chains, for example, reducing them from $N_T$ RF chains to $N_{RF_b}$ for BS, and from $N_t$($N_r$) RF chains to $N_{RF_u}$ for UAV, while satisfying the following conditions: 1) $K \leq N_{RF_b} \ll N_T$; and 2) $K \leq N_{RF_u} \ll N_r$($N_t$). In this case, considering the environment noise follows the distribution  $\mathcal{CN}(\bold{0}, \sigma_n^2)$, the AR for the BS-UAV link can be expressed as follows \cite{ref:23}: 
\begin{equation} \label{eq: 7}
\scalebox{0.9}{$
    \mathrm{R_1}(\bold{F}_b,\bold{B}_b,\bold{F}_{u,r},\bold{B}_{u,r})= \log_2|\bold{I}_K+\bold{Q}_1^{-1}\bold{B}_{u,r}\bm{\mathcal{H}}_1\bold{B}_b\bold{B}_b^H\bm{\mathcal{H}}_1^H\bold{B}_{u,r}^H|$},
\end{equation}
where $\bold{Q}_1^{-1}=(\sigma_n^2\bold{B}_{u,r}\bold{F}_{u,r})^{-1}\bold{F}_{u,r}^H\bold{B}_{u,r}^H$, $\bm{\mathcal{H}}_1=\bold{F}_{u,r}\bold{H}_1\bold{F}_b$, and $\mathbb{E}$\{$\bold{dd}^H$\} $=\bold{I}_K \in \mathbb{C}^{K \times K}$. In the same manner, the AR between UAV and IoT users is calculated based on the instantaneous signal-to-interference-plus-noise ratio (SINR) as demonstrated by the following expression \cite{ref:23}:
\begin{equation} \label{eq: 8}
    \text{SINR}_{g_k} = \scalebox{0.87}{$ \frac{|\bold{h}_{2,k}^H\bold{F}_{u,t}\bold{b}_{u,t, g_k}|^2}{\sum_{\hat{k} \neq k}^{K_g}|\bold{h}_{2,k}^H\bold{F}_{u,t}\bold{b}_{u,t, g_{\hat{k}}}|^2+\sum_{q \neq g}^G\sum_{\hat{k} \neq k}^{K_g}|\bold{h}_{2,k}^H\bold{F}_{u,t}\bold{b}_{u,t, q_{\hat{k}}}|^2 + \sigma_n^2}$}.
\end{equation}
where $g_k = k + \sum_{g'=1}^{g-1} K_{g'}$ represents the IoT user index and $\bold{h}_{2,g_k} \in \mathbb{C}^{N_t}$ is the channel vector between UAV and respective IoT user.
Utilizing the instantaneous SINR, the ergodic AR of the second link $R_2$, in UAV-assisted mmWave MU-mMIMO systems, can be written as:
\begin{equation} \label{eq: 9}
\scalebox{0.9}{$
     \mathrm{R_2}(\bold{F}_{u,t},\bold{B}_{u,t}, x_u, y_u)=\mathbb{E}\left\{\sum_{g=1}^{G}\sum_{k=1}^{K_g}\mathbb{E}\left[\log_2(1+\text{SINR}_{g_k})\right]\right\}$}.
\end{equation}
\vspace{-3em}
\subsection{Channel Model}
We consider the mmWave channel for both links, then using the Saleh-Valenzuela channel model, the channel between BS and UAV can be written as follows \cite{ref:24}:
\begin{equation} \label{eq: 10}
\begin{split}
    \mathbf{H}_1 & = \sum_{c=1}^{C}\sum_{l=1}^{L}z_{1_{cl}}\tau_{l_{cl}}^{-\eta}\mathbf{a}_1^{(r)}(\theta_{cl}^{(r)},\phi_{cl}^{(r)})\mathbf{a}_1^{(t)T}(\theta_{cl}^{(t)},\phi_{cl}^{(t)}) \\
    & = \mathbf{A}_1^{(r)}\mathbf{Z}_1\mathbf{A}_1^{(t)},   
\end{split}
\end{equation}
where $C$ is the total number of groups, $L$ is the total number of paths from BS to UAV, $z_{1_{cl}}\sim\mathcal{CN}(\bold{0}, \frac{1}{L})$ is the complex gain of $l^{th}$ path in the $c^{th}$ cluster, and $\eta$ is the path loss exponent. In addition, $\bold{a}_1^{(k)}(.,.)$ represents the respective transmit or receive array steering vector for a uniform rectangular array (URA), which is defined as \cite{ref:7}:
\begin{equation} \label{eq: 11}
\begin{split}
    & \bold{a}_1^{(k)}(\theta, \phi) = \\
    & [1, e^{-j2\pi d\sin{(\theta)}\cos{(\phi)}}, \cdots, e^{-j2\pi d(N_x-1)\sin{(\theta)}\cos{(\phi)}}] \otimes \\
    & [1, e^{-j2\pi d\sin{(\theta)}\sin{(\phi)}}, \cdots, e^{-j2\pi d(N_y-1)\sin{(\theta)}\sin{(\phi)}}],
\end{split}
\end{equation}
where $k = \left\{r, t\right\}$, $N_x$ and $N_y$ are the horizontal and vertical size of respective antenna array at BS and UAV, $d$ is the inter-element spacing, $\bold{Z}_1$ = diag$(z_{1, 1}\tau_{1, 1}^{-\eta},...,z_{1, L}\tau_{1, L}^{-\eta}) \in \mathbb{C}^{L \times L}$ represents the diagonal gain matrix, $\bold{A}_1^{(r)} \in \mathbb{C}^{N_r \times L}$ and $\bold{A}_1^{(t)} \in \mathbb{C}^{L \times N_t}$ are the receive and transmit phase response matrices, respectively. Also, the angles $\theta_{cl}^{(t)} \in [\theta_c^{(t)} - \delta_c^{\theta(t)}, \theta_c^{(t)} + \delta_c^{\theta(t)}]$ and $\phi_{cl}^{(t)} \in [\phi_c^{(t)} - \delta_c^{\phi(t)}, \phi_c^{(t)} + \delta_c^{\phi(t)}]$ denote the elevation AoD (EAoD) and azimuth AoD (AAoD) for $l^{th}$ path in channel $\bold{H}_1$, respectively. $\theta_c^{(t)}$ represent the mean EAoD and $\delta_c^{\theta(t)}$ is the EAoD spread, while $\phi_c^{(t)}$ is mean AAoD with spread $\delta_c^{\phi(t)}$. In a similar fashion, the angles $\theta_{cl}^{(r)}$ within the range $[\theta_c^{(r)} - \delta_c^{\theta(r)}, \theta_c^{(r)} + \delta_c^{\theta(r)}]$ and $\phi_{cl}^{(r)}$ within $[\phi_c^{(r)} - \delta_c^{\phi(r)}, \phi_c^{(r)} + \delta_c^{\phi(r)}]$ correspond to the elevation AoA (EAoA) and azimuth AoA (AAoA), respectively. Here, $\theta_c^{(r)}$ and $\phi_c^{(r)}$ represent the mean EAoA and AAoA, with $\delta_c^{\theta(r)}$ and $\delta_c^{\phi(r)}$ denoting the angular spreads of the elevation and azimuth angles, respectively. The channel vector between UAV and $k^{th}$ IoT user is written as:
\begin{equation} \label{eq:12}
    \begin{split}
        \mathbf{h}_{2, k}^T = \sum_{q=1}^{Q}z_{2, k_q}\tau_{2, k_q}^{-\eta}\bold{a}(\theta_{k_q}, \phi_{k_q})  = \mathbf{z}_{2, k}^{T}\mathbf{A}_{2,k} \in \mathbb{C}^{N_t}, 
    \end{split}
\end{equation}
where $Q$ is the total number of downlink paths from UAV to users, $z_{2,k_q} \sim \mathcal{CN}(0, \frac{1}{Q})$ is the complex path gain of $q^{th}$ path in the second link, and $\bold{a}(.,.) \in \mathbb{C}^{N_t}$ is the UAV downlink array phase response vector. As given in (\ref{eq:12}), the obtained downlink channel is comprised of two distinct parts: 1) a fast time-varying path gain vector $z_{2,k} = [z_{2,k_1}\tau_{2, k_1}^{-\eta}, ..., z_{2,k_Q}\tau_{2, k_Q}^{-\eta}]^T \in \mathbb{C}^Q$; and 2) a slow time-varying downlink array phase response matrix $\bold{A}_{2,k} \in \mathbb{C}^{Q \times N_t}$ where each row is constituted by $\bold{a}(\theta_{kl}, \phi_{kl})$. Afterward, the channel matrix for the second link can be expressed as follows:
\begin{equation} \label{eq: 13}
    \mathbf{H}_2 = [\mathbf{h}_{2, 1}, \cdots, \mathbf{h}_{2,K}]^T = \mathbf{Z}_2\mathbf{A}_2 \in \mathbb{C}^{K \times N_t},
\end{equation}
where $\mathbf{Z}_2 = [\mathbf{z}_{2,1}, \cdots, \mathbf{z}_{2, K}]^T \in \mathbb{C}^{K \times Q}$ is the complete path gain matrix for all downlink IoT users.
\vspace{-1ex}
\section{Joint HBF \& DDPG-Based UAV Deployment}
\vspace{-1ex}
In this section, our objective is to jointly optimize the UAV location and HBF for BS and UAV to reduce the channel state information (CSI) overhead size while maximizing the total AR of UAV-assisted MU-mMIMO IoT systems. First, we design the stages $\mathbf{F}_b, \mathbf{F}_{u,r}, \mathbf{F}_{u,t}$ based on the slow time-varying AoD and AoA. Then, the BB stages $\mathbf{B}_b, \mathbf{B}_{u,r}, \mathbf{B}_{u,t}$ are developed by using singular value decomposition (SVD).
\vspace{-1ex}
\subsection{HBF Design}
The RF and BB stages for BS and UAV are designed for the following considerations: 1) maximize the beamforming gain at the desired directions based on the slow time-varying AoD and AoA; 2) reduce the power-hungry RF chains; 3) reduce the CSI overhead; and 4) mitigate multi-user interference (MU-I)\footnote{The details for HBF design for BS and UAV can be found in \cite{ref:23}}. 
\subsection{DDPG: Preliminaries}
DDPG is a sophisticated RL algorithm that combines elements of deep Q-networks (DQN) and policy gradient techniques. Unlike DQN, which is designed for discrete action spaces, DDPG is tailored for continuous action spaces, common in real-world situations. This makes DDPG ideal for complex tasks that demand a spectrum of continuous action values, increasing its effectiveness in diverse and changing environments. \par
DDPG utilizes actor-critic approach \cite{ref:19}, where the actor, denoted as $\mu(\mathbf{s}|\theta^\mu)$, outputs a deterministic action $\mathbf{a}$ given a state $\mathbf{s}$, where $\theta^\mu$ represents the weights of the actor network. In the same manner, critic expressed as $Q(\mathbf{s},\mathbf{a}|\theta^Q)$, predicts the expected return (value) of taking an action $\mathbf{a}$ in a state $\mathbf{s}$, and it is parameterized by weights $\theta^Q$. DDPG uses target networks for both actor and critic, represented as $\mu'(\mathbf{s}|\theta^{\mu'})$ and $Q'(\mathbf{s},\mathbf{a}|\theta^{Q'})$ respectively. Here, $\theta^{\mu'}$ and $\theta^{Q'}$ denote the weights of the target actor and target critic networks. These target networks are essentially slow-paced versions of the primary actor and critic networks, ensuring a more stable and consistent learning process. Moreover, to break the correlation between consecutive experiences ($\mathbf{s}_t$, $\mathbf{a}_t$, $r_t$, $\mathbf{s}_{t+1}$), we use a replay buffer $D$, where $\mathbf{s}_t$, $\mathbf{a}_t$, $r_t$ are the state, action, and reward at timestep $t$, respectively, and $\mathbf{s}_{t+1}$ is the next state. These transitions are stored in the replay buffer, and then at each timestep, the actor and critic are updated by sampling a minibatch of transitions $(\mathbf{s}_i, \mathbf{a}_i, r_i, \mathbf{s}_{i+1})$ uniformly from the buffer. During the training, the weights of the critic are updated by minimizing the mean square error (MSE) loss function as:
\begin{equation} \label{eq: 33}
    \mathcal{L} = \frac{1}{N}\sum_{i}(y_i-Q(\mathbf{s}_i, \mathbf{a}_i|\theta^Q))^2,
\end{equation} 
 where $N$ is the batchsize and $y_i$ is defined as:
 \begin{equation} \label{eq: 34}
     y_i = r_i + \gamma Q'(\mathbf{s}_{i+1}, \mu'(\mathbf{s}_{i+1}|\theta^{\mu '})|\theta^{Q'}).
 \end{equation}
 Here, $\gamma \in [0, 1]$ represents the discounting factor. Then, the actor network is updated using the policy gradient method as:
 \begin{equation} \label{eq: 35}
     \nabla_{\theta^\mu}J \approx \frac{1}{N} \sum_i\nabla_aQ(\mathbf{s},\mathbf{a}|\theta^Q)|_{\mathbf{s}=\mathbf{s}_i, a=\mu_{\mathbf{s}_i}}\nabla_{\theta^\mu}\mu(\mathbf{s}|\theta^\mu)|_{\mathbf{s}_i}.
 \end{equation}
 Afterward, the target actor and target critic networks are updated using the soft update approach:
 \begin{equation} \label{eq: 36}
     \theta^{Q'} \leftarrow \tau\theta^Q + (1-\tau)\theta^{Q'}
 \end{equation}
 \begin{equation} \label{eq: 37}
     \theta^{\mu'} \leftarrow \tau\theta^\mu + (1-\tau)\theta^{\mu'}
 \end{equation}
 where $\tau$ is a small coefficient denoting how fast the target actor and critic are updated. As in every RL algorithm, we need exploration to achieve the best policy. In this regard, we add a zero-mean Gaussian noise $\mathcal{N}$ to the actor policy as follows:
 \begin{equation} \label{eq: 38}
     \mathbf{a}_t = \mu(\mathbf{s}_t|\theta_t^\mu)+\mathcal{N},
 \end{equation}
 where $\mathcal{N}$ follows the distribution $\mathcal{CN}(0, \sigma^2)$. 

\begin{algorithm}
\label{alg:ddpg}
\caption{DDPG-Based UAV Deployment}
\SetAlgoLined
Randomly initialize actor $\mu(\mathbf{s}|\theta^{\mu})$ and critic $Q(\mathbf{s},\mathbf{a}|\theta^Q)$ networks with weights $\theta^\mu$ and $\theta^Q$.\\
 Initialize target network $Q'$ and $\mu'$ with weights $\theta^{Q'} \leftarrow \theta^Q$, $\theta^{\mu'} \leftarrow \theta^\mu$. \\
 Initialize replay buffer $D$.\\
 \For{episode=1:M}{
  Receive initial observation state $\mathbf{s}_1$. \\
  \For{t=1:T}{
   Select action $\mathbf{a}_t$ using (\ref{eq: 38}).\\
   Execute action $\mathbf{a}_t$ and observe reward $r_t$ and new state $\mathbf{s}_{t+1}$.\\
   Store transition $(\mathbf{s}_t, \mathbf{a}_t, r_t, \mathbf{s}_{t+1})$ in $D$.\\
   Sample a random minibatch of $N$ transitions $(\mathbf{s}_i, \mathbf{a}_i, r_i, \mathbf{s}_{i+1})$ from $D$.\\
   Formulate $y_i$ using (\ref{eq: 34}).\\
   Update the critic by minimizing the loss function $\mathcal{L}$ in (\ref{eq: 33}).\\
   Update the actor using the policy gradient in (\ref{eq: 35}).\\
   Update the target networks via (\ref{eq: 36}), (\ref{eq: 37}).
   }
}
\end{algorithm}
\vspace{-1em}
\subsection{DDPG-Based UAV Deployment}
To apply the DDPG algorithm to our problem, we need to define appropriate states, actions, and rewards for the agent. Moreover, the design of the actor and critic network can significantly influence the performance of the algorithm. Thus, first, we introduce the suitable state, action, and reward for UAV deployment, and then, we discuss the configuration of the actor and critic network.
\subsubsection{States} At each time step $t$, the UAV agent will observe state  $\mathbf{s}_t$ as follows:
\begin{equation} \label{eq: 39}
    \mathbf{s}_t = [\hat{x}_{u,t}, \hat{y}_{u,t}]^T \in \mathbb{R}^2,
\end{equation}
where $\hat{x}_{u,t}=\frac{x_{u,t}}{x_{\text{max}}}$ and $\hat{y_{u,t}} = \frac{y_{u,t}} {y_{\text{max}}}$ denote the 2D normalized location of the UAV at time step $t$. Here, we assume the UAV is deployed at a fixed height $z_{u,t}$\footnote{For simplicity, we consider a scenario of fixed UAV height. However, the proposed DDPG-based solution can be applied for 3D UAV deployment, which is left as our future work.}. 
\subsubsection{Actions} We consider action $\mathbf{a}_t$ at time step $t$ for the UAV  agent as follows:
\begin{equation} \label{eq: 40}
    \mathbf{a}_t = [a_{t,x}, a_{t,y}]^T \in \mathbb{R}^2 ,  \quad  \left\{ \begin{array}{rcl}
a_{t,x} \in [-a_{x, \text{max}}, a_{x, \text{max}}] \\
a_{t,y} \in [-a_{y, \text{max}}, a_{y, \text{max}}]
\end{array}\right.
\end{equation} 
Here, $a_{x, \text{max}}$ ($a_{y, \text{max}}$) represents the maximum movement step for the UAV on the x-axis (y-axis). When $a_{t,x}$ ($a_{t,y}$) is positive, the movement is to the East (North), and when $a_{t,x}$ ($a_{t,y}$) is negative, the movement is to the West (South). 
\subsubsection{Reward}The reward function effectively communicates the objectives of the task to the agent. Designing the reward function correctly is pivotal since it not only shapes the learning trajectory but also influences the convergence speed and the overall effectiveness of the policy learned.  Considering this, we define the reward function $r_t$ at the time step $t$ as:
\begin{equation} \label{eq: 41}
    r_t = \left\{ \begin{array}{rcl}
\mathrm{R_2} & \mbox{for} & \mathrm{R_2} \geq \eta_0 \\
-1 & \mbox{for} & \mathrm{R_2} < \eta_0 \\
-5 & \mbox{for} & x_{u,t} > x_{\text{max}}\ \text{or}\ y_{u,t} > y_{\text{max}} \\
-5 & \mbox{for} & x_{u,t} < x_{\text{min}}\ \text{or}\ y_{u,t} < y_{\text{min}}\\
\end{array}\right.
\end{equation}
where $\mathrm{R_2}$ denotes the achievable rate for the second link as given in (\ref{eq: 9}), $\eta_0$ is an adjustable threshold,  $x_{\text{max}}$ ($x_{\text{min}}$) is the maximum (minimum) allowable position on the x-axis, and $y_{\text{max}}$ ($y_{\text{min}}$) is the maximum (minimum) permitted position on the y-axis.
\subsubsection{Actor and Critic Networks}
We employ a fully connected deep neural network (DNN) architecture with two hidden layers as depicted in Fig. 2, for both (target) actor and (target) critic networks. Here, the actor network predicts suitable actions as described in (\ref{eq: 40}) based on the input states given in (\ref{eq: 39}), whereas, the critic network determines the Q-value of the input action-state pair. We consider $L_i^a$ neurons in each $i^{th}$ hidden layer for the actor network with $i = \{1, 2\}$. Similarly, we will have $L_j^c$ neurons in each $j^{th}$ hidden layer for the critic network with $j = $\{1,2\}. \par
\setlength{\belowcaptionskip}{2pt}
\begin{table}[!t]
\caption{Simulation Parameters}
\scalebox{0.69}{
\begin{tabular}{|c|c|c|c|}
\hline
\multicolumn{2}{|c|}{Number of antennas} & \multicolumn{2}{|c|}{($N_T, N_t, N_r$) = 144} \\ \hline
BS height & UAV height & 10 m & 20 m \\ \hline
UAV x-axis range & UAV y-axis range & [$x_{\text{min}}, x_{\text{max}}$] = [0,100]m & [$y_{\text{min}}, y_{\text{max}}$] = [0,100]m \\ \hline
UAV x-axis movement & UAV y-axis movement & [$a_{x,\text{min}}, a_{x,\text{max}}$] = [-1,1]m & [$a_{y,\text{min}}, a_{y,\text{max}}$] = [-1,1]m \\ \hline
User groups & \# of users per group & $G=1$ & $K_g = \frac{K}{G}$ \\ \hline
\# of paths & Path loss exponent & $L=10$ & 3.6 \\ \hline
Noise PSD & Reference path loss $\alpha$ & -174 dBm/Hz & 61.34 dB \\ \hline
Frequency & Channel bandwidth & 28 GHz & 100 MHz \\ \hline
Mean AAoD/AAoA ($1^{st}$ link) & Mean AAoD ($2^{nd}$ link) & $120^{\circ}$ & $\phi_g = 21^{\circ}+120^{\circ}(g-1)$ \\ \hline
Mean EAoD/EAoA ($1^{st}$ link) & Mean EAoD ($2^{nd}$ link) & $60^{\circ}$ & $\theta_g = 60^{\circ}$ \\ \hline
Azimuth/Elevation angle spread & \# of network realization & $\pm10^{\circ}$ & 2000 \\ \hline
\end{tabular}}
\label{table 1}
\end{table}
\begin{figure}[!t]
    \centering
    \includegraphics[width=\linewidth]{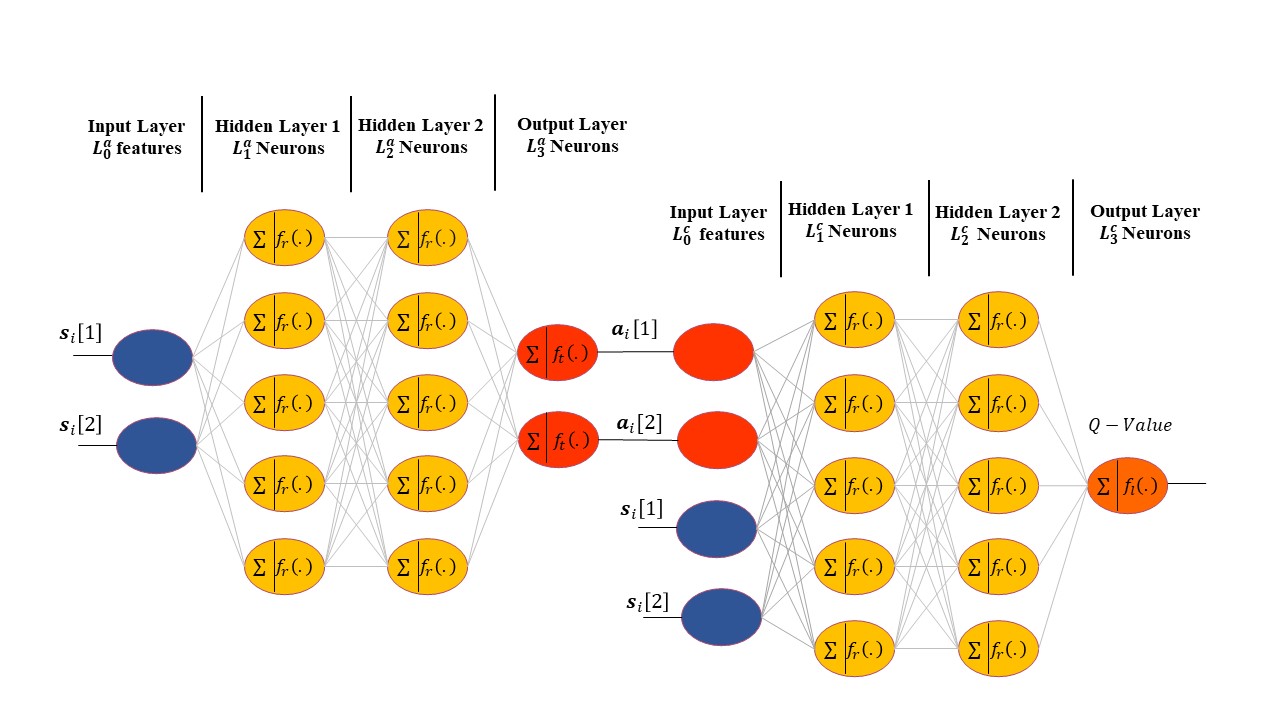}
    \caption{(Target) actor and (target) critic DNN architecture}
    \label{fig:Actor-Critic Architecture}
    \vspace{-2em}
\end{figure}
To perform non-linear operations, we utilize the rectified linear unit (ReLU) as the activation function in the hidden layers for both actor and critic networks (i.e., $f_r(z) = \text{max}(0,z)$). To ensure that the predicted actions by the actor network are between $[-a_{x, \text{max}}, a_{x, \text{max}}]$ ($[-a_{y, \text{max}}, a_{y, \text{max}}]$), we must use a function that can output both negative and positive values. Thus, we apply \textit{tanh} activation function in the output layer of the actor network (i.e., $f_t(z)=\frac{e^z-e^{-z}}{e^z+e^{-z}}$). However, since critic network estimates the Q-value function, the range of outputs is quite large or unbounded. Therefore, we use the linear activation function for the output layer of the critic network (i.e., $f_l(z) = z$). The summary of the DDPG algorithm is outlined in Algorithm 1. 
\vspace{-1em}
\section{Illustrative results}
In this section, we present illustrative results on the performance of the DDPG-based UAV deployment (DDPG-UD) algorithm for different scenarios. For benchmark comparison, we compare the proposed DDPG-UD with the following solutions: 1) particle swarm optimization (PSO)-based UAV deployment (PSO-UD); and 2) deep learning (DL)-based UAV deployment (DL-UD), (i.e., supervised learning (SL) approach \cite{ref:23}). Table I outlines the simulation setup based on the 3D micro-cell scenario \cite{ref:7}, whereas the hyper-parameters of the DDPG algorithm are given in Table II. In the following, we compare the performance for both static (fixed user locations) and dynamic (users changing locations) environments in MU-mMIMO IoT systems.
\setlength{\belowcaptionskip}{2pt}
\begin{table}[!t]
\centering
\caption{Networks' Parameters}
\scriptsize
\begin{tabular}{|c|c|c|c|}
\hline
\multicolumn{4}{|c|}{(Target) Actor Network Architecture} \\ \hline 
Input Shape & $L_0^a = 2$ & $1^{st}$ hidden layer & $L_1^a = 20$ \\ \hline 
$2^{nd}$ hidden layer& $L_2^a = 20$ & Output layer &  $L_3^a = 2$ \\ \hline 
\multicolumn{4}{|c|}{(Target) Critic Network Architecture} \\ \hline
Input Shape & $L_0^c = 4$ & $1^{st}$ hidden layer & $L_1^c = 20$ \\ \hline
$2^{nd}$ hidden layer & $L_2^c = 20$ & Output layer & $L_3^c = 1$ \\ \hline
\multicolumn{4}{|c|}{Network Parameters} \\ \hline
Reply buffer size & 60000 & Critic learning rate & 0.002 \\ \hline
Actor learning rate & 0.001 & Target networks learning rate & 0.01 \\ \hline
\end{tabular}
\label{table 2}
\vspace{-1em}
\end{table} 

\subsection{Static Environment (Fixed Users Location)}
\indent In this section, we consider IoT users to have a fixed location and discuss two scenarios: 1) narrow-range user distribution; and 2) wide-range user distribution. For narrow-range user distribution, we consider that the BS is located at $(x_b,y_b,z_b) = (0,0,10)$, the UAV initial position is $(x_u,y_u,z_u) = (50,50,20)$, and $K=4$ IoT users are distributed randomly at a far distance from the BS (i.e., $(x_k,y_k) \in [90,100]$). UAV starts its initial location at $(50,50,20)$ at the beginning of each episode, and then it explores the environment during the time steps. For wide-range user distribution, we consider the same location and initial position for the UAV, while assuming that $K=4$ IoT users are randomly scattered with $(x_k,y_k) \in [50,100]$. Fig. 3 shows the achieved rates for DDPG-UD, PSO-UD, DL-UD, and fixed deployment (FD) (i.e., no optimization). Numerical results reveal that the proposed DDPG-UD can achieve 98.63\% of PSO-UD performance for narrow-range user distribution while having a 16 times better performance than FD. Furthermore, DDPG-UD can enhance the performance of PSO-UD by 2.23\% for wide-range user distribution while having 3.14 times better AR than the FD. Fig. 3 also demonstrates that although DL-UD can have an acceptable performance when we have narrow-range user distribution while achieving 89.62\% AR of PSO-UD, it fails to find the optimal location for wide-range user distribution having only 36.94\% AR of the DDPG-UD. This means that for more complex user distributions, DL-UD is not a promising solution. For the next step, we consider that IoT users may change their location during the observation.
\vspace{-1em}
\subsection{Dynamic Environment (Changing Users Locations)}
In this section, we consider a more practical scenario where IoT users can change their location during the training. This scenario represents dynamic, real-world environments. We assume that the BS is set at $(x_b,y_b,z_b) = [0,0,10]$. Then we set the IoT users' locations randomly at six different distributions $l \in$ \{$l_1, l_2, ..., l_6$\} as follows:
\begin{equation}
(x_{k,l}, y_{k,l})\!=\! 
    \left\{ \begin{array}{rcl}
x_{k,l} \in [60, 70], \; y_{k,l} \in [60, 70] & \mbox{for} & l_1 \\
x_{k,l} \in [60, 70], \; y_{k,l} \in [70, 80] & \mbox{for} & l_2 \\
x_{k,l} \in [70, 80], \; y_{k,l} \in [80, 90] & \mbox{for} & l_3 \\
x_{k,l} \in [80, 90], \; y_{k,l} \in [80, 90] & \mbox{for} & l_4 \\
x_{k,l} \in [80, 90], \; y_{k,l} \in [70, 80] & \mbox{for} & l_5 \\
x_{k,l} \in [80, 90], \; y_{k,l} \in [60, 70] & \mbox{for} & l_6 \\
\end{array}\right.
\end{equation}
where $x_{k,l}$ and $y_{k,l}$ denote the $k^{th}$ IoT user x-axis and y-axis range during $l^{th}$ distribution, respectively. Furthermore, UAV starts its initial location at $(x_u, y_u, z_u) = (50, 50, 20)$ and finds the optimal deployment $\mathbf{x}_o^{(1)} = \{x_o^{(1)}, y_o^{(1)}\}$ for $l_1$, which is now used as its initial location for the next user distribution (i.e., $l_2$). Fig. 4 shows the accumulated reward and average accumulated reward for the DDPG agent during six different user locations. This figure shows that after the first user location, the DDPG agent maintains the reward and finds the optimal location in less number of episodes. The reason is that the pre-trained networks from previous user locations contain information about the environment, thus helping the DDPG agent find the optimal UAV location for changing user locations in a shorter time.
Fig. 5 shows the achieved rate for DDPG-UD, PSO-UD, DL-UD, and FD, which shows that DDPG-UD can achieve a performance close to PSO-UD. For instance, DDPG-UD accomplishes 99.62\% of PSO-UD AR for $l_3$, and it provides 18.54 bps/Hz AR at $l_6$ and achieves 99.42\% of PSO-UD performance. Additionally, the capacity is improved by 36.99\% and 16.73\% for FD for $l_3$ and $l_6$, respectively. Fig. 5 also supports the previous deduction that DL-UD is not a promising solution for complex scenarios while having only 63.84\% and 66.65\% of PSO-UD performance for $l_3$ and $l_6$, respectively. 
\begin{figure}[!t]
    \centering
    \includegraphics[width=\linewidth]{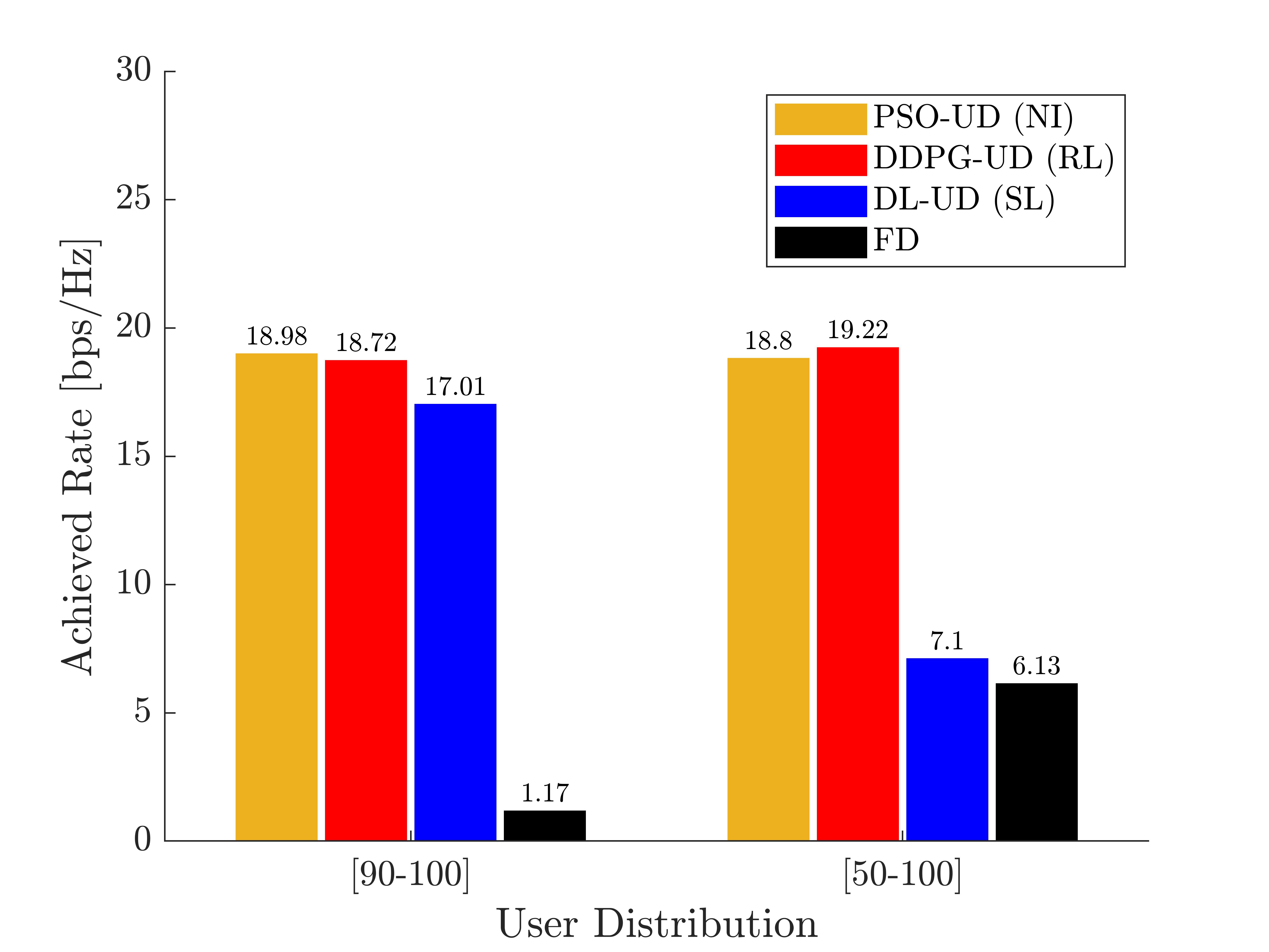}
    \caption{Achieved Rates versus user distribution}
    \label{fig:Single Trajectory}
    \vspace{-1ex}
\end{figure}
\begin{figure}[!t]
    \centering
    \includegraphics[width=\linewidth]{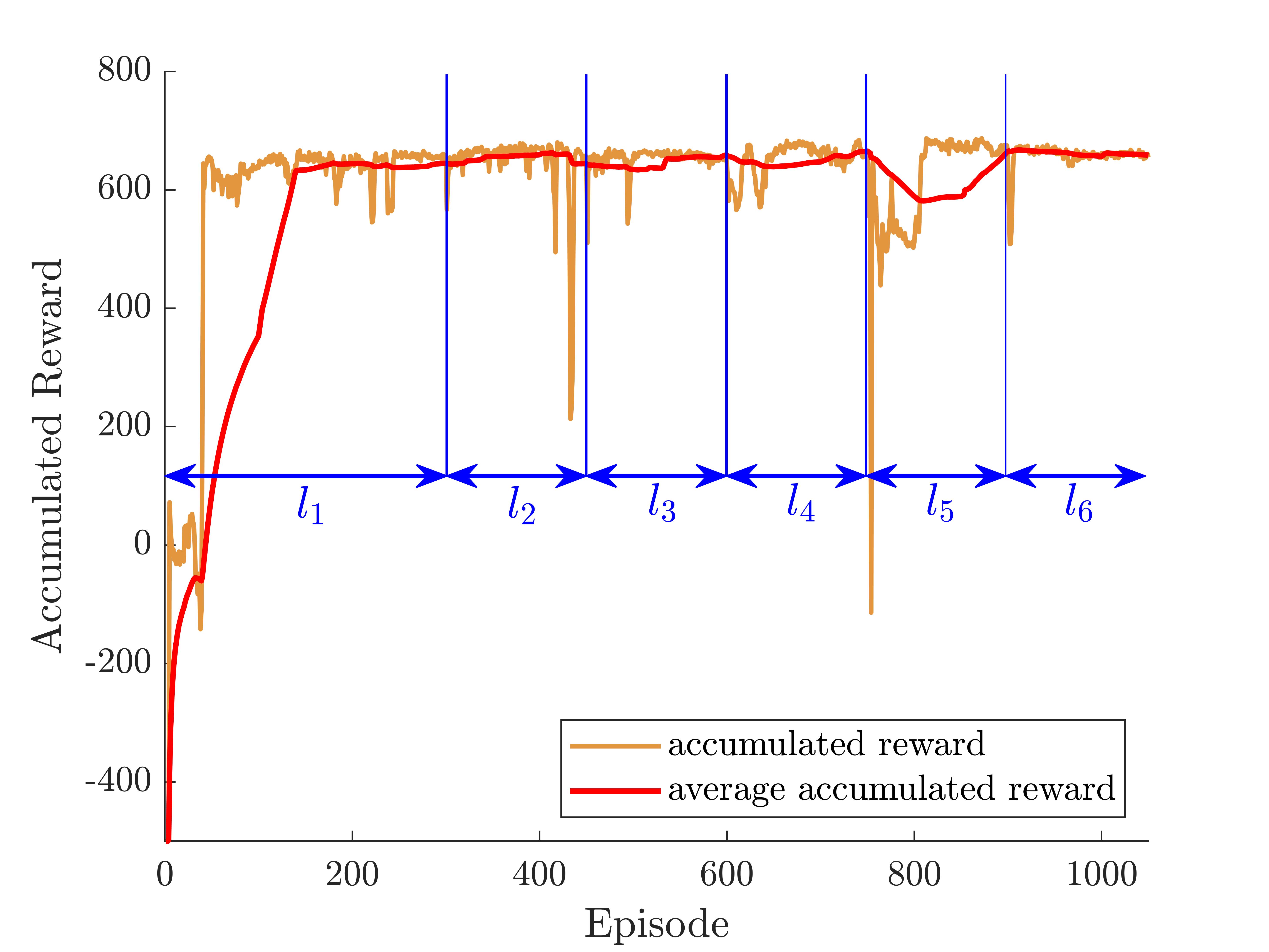}
    \caption{Accumulated reward vs episode for DDPG}
    \label{fig:Score}
    \vspace{-2em}
\end{figure}

\begin{figure}[!t]
    \centering
    \includegraphics[width=\linewidth]{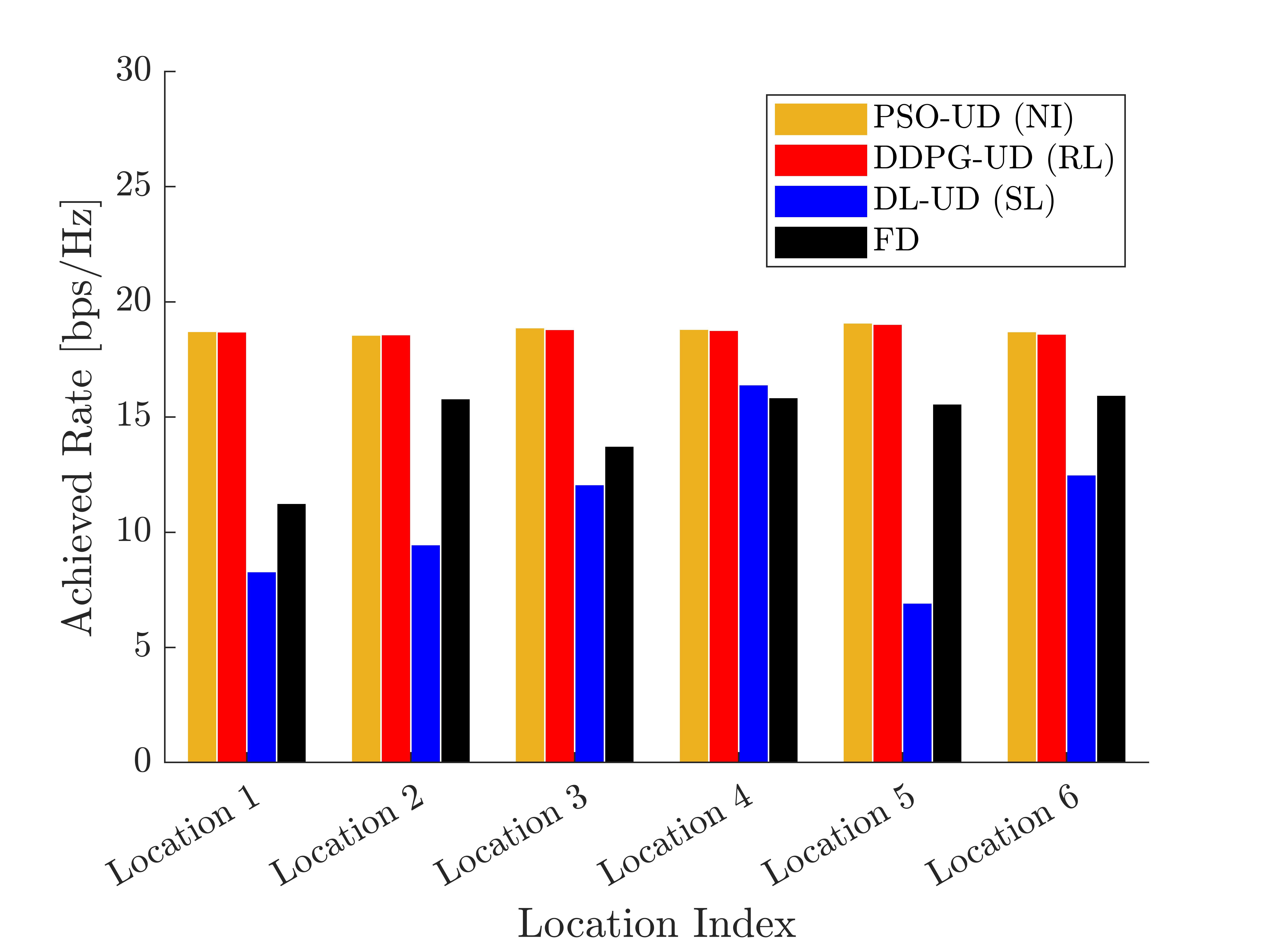}
    \caption{Achieved rates in different locations}
    \label{fig:Dynamic_Comparison}
    \vspace{-1ex}
\end{figure}
Fig. 6 displays the time comparison between DDPG-UD and PSO-UD. Here, we provide the runtime for a single user location (i.e., $l \in l_1$), three different user locations ($l \in \{l_1,l_2,l_3\}$), and six distinct user locations ($l \in \{l_1,\cdots,l_6\}$). This figure shows that although for a single user location, DDPG-UD takes 18.04\% more than PSO-UD to find the optimal location, for multiple user locations, it takes 76.45\% and 68.50\% of PSO-UD runtime. This means that as the number of user locations is increased, DDPG-UD will take less time to find the optimal location when compared to PSO-UD, which makes DDPG-UD an efficient solution for dynamic and fast-changing environments in MU-mMIMO IoT systems.
\vspace{-1em}
\section{Conclusion}
In this work, a novel DDPG-based UAV deployment (DDPG-UD) and hybrid beamforming (HBF) technique has been proposed for AR maximization in dynamic MU-mMIMO IoT systems. First, we introduce HBF for both base station (BS) and UAV. Afterward, we apply the deep deterministic policy gradient (DDPG) as a reinforcement learning approach to UAV deployment in dynamic environments. Illustrative results show that the proposed DDPG-UD closely approaches the optimal rate achieved by particle swarm optimization (PSO)-based UAV deployment (PSO-UD). On the other
hand, DDPG-UD greatly reduces the runtime by 31.5\%, which makes DDPG-UD a more appropriate solution for real-world dynamic applications in UAV-assisted mMIMO IoT systems.
\begin{figure}[!t]
    \centering
    \includegraphics[width=\linewidth]{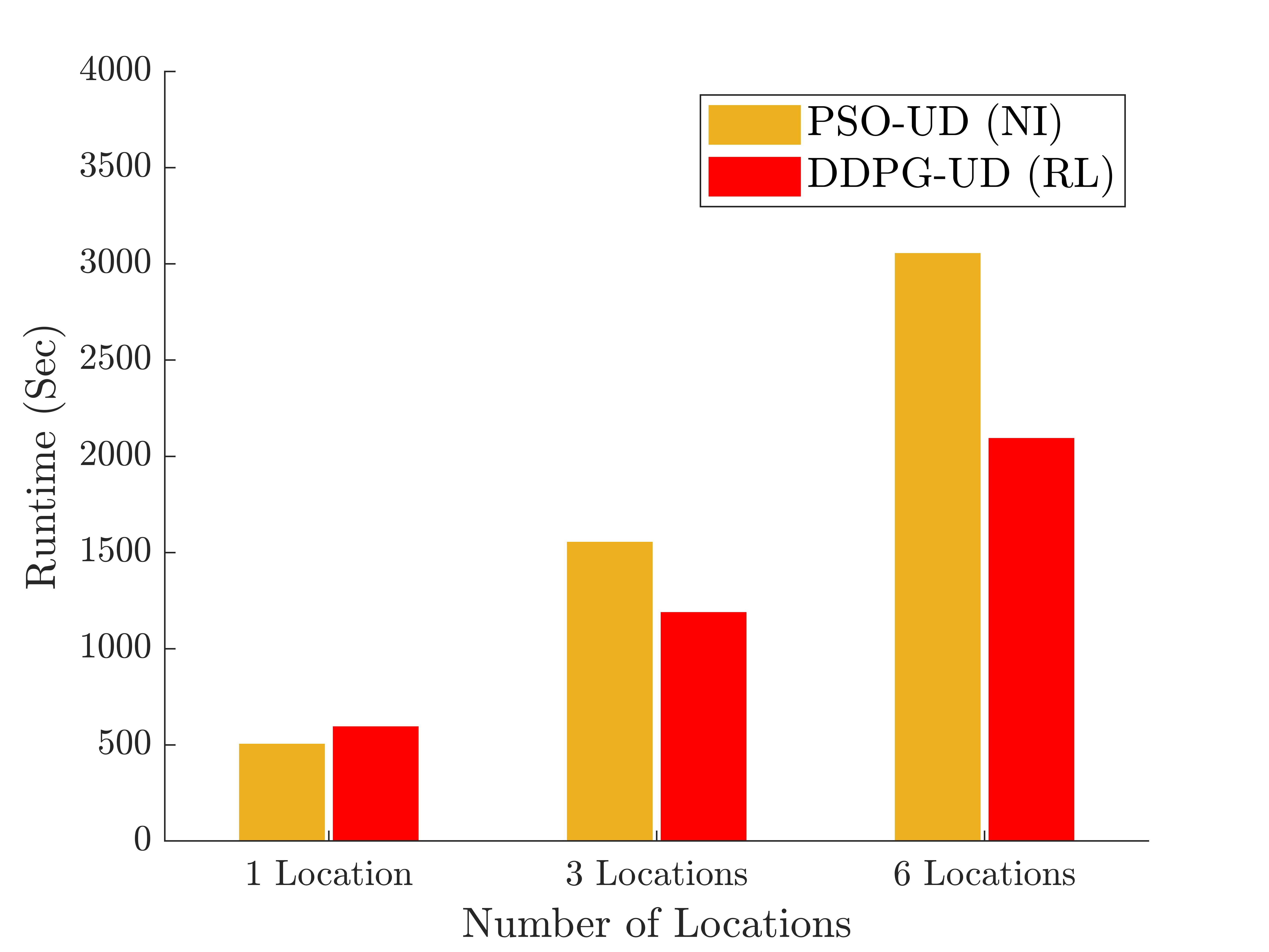}
    \caption{Runtime for different numbers of locations}
    \label{fig:Time}
    \vspace{-1em}
\end{figure}





%

\end{document}